\documentstyle{article}
\begin{document}
\fontsize{10 pt}{13 pt}
\selectfont
\begin{center}
{\bf ENERGY OF THE TAUB COSMOLOGICAL SOLUTION} \vspace{1 cm}

\fontsize{8pt}{10pt} \selectfont PAUL HALPERN \\
\vspace{2 pt}
{\it Department of Mathematics, Physics and Computer Science,\\
University of the Sciences in Philadelphia, 600 S. 43rd St.,\\
Philadelphia, PA. 19104, USA\\ p.halper@usip.edu}
\end{center}
\begin{quote}
\fontsize{8pt}{10pt} \selectfont We calculate the total energy of
Taub's 1951 exact solution for a Bianchi type-IX geometry using
several different energy-localization procedures, including the
prescriptions of Einstein, Papapetrou, Landau-Lifshitz and
M{\o}ller. We compare these results to those for other anisotropic
geometries, and comment on their relationship to Rosen's conjecture
about the total energy of the universe.


\end{quote}
\fontsize{10 pt}{13 pt} \selectfont \vspace{1 cm} \noindent
 {\bf 1. Introduction}

\vspace{6 pt} A longstanding question in general relativity is
whether gravitational energy can be localized and, if so, what
measure to use.  Einstein proposed the first energy-momentum complex
in an attempt to define the local distribution of energy and
momentum \cite{einstein}.  Einstein's measure is non-tensorial and
asymmetric in its indices.  These issues have inspired a number of
alternative measures, including prescriptions by M{\o}ller
\cite{moller}, Tolman \cite{tolman}, Papapetrou \cite{papapetrou},
Landau and Lifshitz \cite{landau}, Weinberg \cite{weinberg} and so
forth. By in large, these formulations depend on the
coordinate-system used. For example, Einstein's prescription favors
quasi-Cartesian coordinates, defined as those coordinates $x^i$ for
which the metric components $g_{\mu\nu}$ converge sufficiently rapidly
 toward the constant components of the diagonal Minkowski metric
  ${\eta}_{\mu\nu}$. Some of these complexes have other
restrictions.

Such limitations have led some researchers to discount the
importance of localized measures. However, in 1990 Bondi proved that
energy must be localizable, at least in principle  \cite{bondi}. The
same year, Virbhadra demonstrated that these complexes, though
non-tensors, yield credible, consistent results when applied to the
same spacetime geometries \cite{virbhadra1}.  In 1993, Rosen and
Virbhadra employed the Einstein prescription and found reasonable
results for the energy and momentum of cylindrical gravitational
waves \cite{rosenvirb}. Later, Virbhadra derived similar results
using Tolman's procedure and the measure proposed by Landau and
Lifshitz \cite{virbhadra2}.  Then in 1996, Aguirregabiria, Chamorro
and Virbhadra found that for any metric of the Kerr-Schild class
(including the Schwarzschild, Reissner-Nordstr{\"o}m, Kerr and
Vaidya metrics, for example.), various energy-momentum prescriptions
yielded reasonable and consistent results \cite{acv}. Of these,
Virbhadra found that the Einstein prescription
offered the greatest consistency \cite{virbhadra3}. These findings
have stimulated a number of investigations attempting to extend such
methods to other classes of metrics \cite{xulu}.

Much work has been done, for example, in using various measures to
determine the energy distributions of assorted generalizations of
black holes. Chamorro and Virbhardra \cite{chamorro} as well as Xulu
\cite{xulu1} have examined the energy distribution of charged black
holes with a dilaton field. Radinschi and Yang \cite {radinschi},
Vagenas \cite{vagenas1}, Gad \cite{gad1} and Xulu \cite{xulu2} have
each looked at the energy distribution of stringy black holes.
Vagenas has also examined black holes in 2+1 dimensions
\cite{vagenas2, vagenas3} and applied M{\o}ller's prescription
to the dyadosphere of a Reissner-Nordstr{\"o}m black hole \cite{vagenas4}.
The energy distribution of black plane solutions
has also been investigated \cite{halpern}. Other researchers have
looked at the energy distribution for the case of teleparallel
gravity \cite{andrade}, for a stationary beam of light
\cite{bringley,gad2}, for a radiating charged particle \cite{vagenas5}
and for a host of other metrics.

Cosmological solutions are another area of considerable interest,
stimulated by a conjecture made by Rosen in 1994 that the total
energy of the universe, presuming it has a closed geometry, is zero
\cite{rosen}.  Applying the Einstein energy-momentum complex to a
closed Friedmann-Robertson-Walker (FRW) cosmology, Rosen determined
its total energy to be indeed zero.  Johri, Kalligas, Singh and
Everitt later reached similar conclusions by applying the
prescription of Landau and Lifshitz \cite{johri}.

One possible implication of Rosen's conjecture is a hypothesis
proposed by Cooperstock that the energy distribution is non-zero
only in regions where the stress-energy tensor $T_{\mu \nu}$ is
non-zero \cite{cooperstock}. This implies that gravitational waves
traveling through a vacuum would carry only information content, not
energy, and thus would be fundamentally different from other forms
of radiation.

Researchers have sought to extend these results to more general
cosmologies. The isotropic FRW model is far from the most general
cosmological solution, and therefore would not, for example,
represent a chaotic very early stage of the universe.  The simplest
anisotropic model is the Kasner solution, possessing a Bianchi type
I geometry. Several researchers have investigated the total energy
of this model using a variety of measures, and have found the result
to be identically zero for any closed region, consistent with
Rosen's conjecture \cite{banarjee, radinschi2, xulu3, aydogdu,
salti}.  In addition to the standard vacuum case, Salti looked at
the viscous Kasner universe in teleparallel gravity \cite{salti2}
using the Einstein and Landau-Lifshitz formulations and found the
total energy and momentum to be identically zero. Generalizing these
findings to another anisotropic cosmology,
Radinschi examined the energy-momentum of the Bianchi type VIO
universe using a number of different measures, finding that all the
components vanish \cite{radinschi3}.

\fontsize{10 pt}{13 pt} \selectfont \vspace{1 cm} \noindent
 {\bf 2. The Taub Cosmological Solution}

\vspace{6 pt}
Amongst homogenous cosmologies, the metric that yields
the most complex anisotropic behavior possesses a Bianchi type IX
spatial geometry.  The one-forms associated with the spatial part of
this metric possess the Lie algebra associated with one of the
normal subgroups of the group of motion of a three-dimensional
sphere.

As Belinski, Khalatnikov and Lifshitz demonstrated in 1970, in
evolving the vacuum case backward in time toward its initial
singularity, the behavior of its three scale factors resembles a
sequence of transitions between Kasner-like epochs \cite{belinsky}.
In using Hamiltonian methods to analyze this behavior, Misner dubbed
it the ``Mixmaster universe,'' because of its special mixing
properties \cite{misner}. Barrow later demonstrated that the vacuum
Bianchi type IX solution exhibits a form of deterministic chaos
\cite{barrow}.

If one constrains two of the three scale factors to be equal, one
obtains an exact, non-chaotic solution,  first identified by Taub in
1951.  The Taub metric can be expressed in a holonomic coordinate
system as:

\begin{eqnarray}
ds^2 &=& \gamma(t)\hspace{0.1cm} dt^2 - R(t)\hspace{0.1cm}{dx}^2
- ( R(t)\hspace{0.1cm} \sin^2 (x) + S(t)\hspace{0.1cm} \cos^2 (x) ) \hspace{0.1cm}{dy}^2 \nonumber \\
&-& 2 \hspace{0.1cm}S(t) \hspace{0.1cm} \cos(x)\hspace{0.1cm} dy
\hspace{0.1cm}dz - S(t) \hspace{0.1cm} {dz}^2
\end{eqnarray}

with $x$, $y$ and $z$ corresponding to the Euler angles
of three-dimensional rotation, and having the ranges:

\begin{equation}
0 \le x \le \pi ; \hspace{0.3cm} 0 \le y \le 4 \pi ; \hspace{0.3cm}
0 \le z \le 2 \pi
\end{equation}

As Taub demonstrated, to satisfy the vacuum Einstein equations
 the scale factors are constrained to be:

\begin{eqnarray}
R(t) &=& \frac{k \hspace{0.1cm} \cosh(k \hspace{0.1cm} t + \alpha)}
{4 \hspace{0.1cm} \cosh^2(\frac{k \hspace{0.1cm} t + \beta}{2})}\\
S(t) &=& \frac{k}{\cosh(k \hspace{0.1cm} t + \alpha)}
\end{eqnarray}

where $\alpha$, $\beta$ and $k$ are constants and with:

\begin{equation}
\gamma(t) = {R(t)}^2 S(t)
\end{equation}

One reason the Taub solution is of interest is that in 1968, Misner
and Taub \cite{misner2} proved that it represents the
spacetime region bounded by the horizon of the NUT solution found by
Newman, Unti and Tamburino \cite{newman}.  Joined together, these
constitute the Taub-NUT spacetime.

\vspace{1 cm} \noindent {\bf 3. Finding the Energy Using Einstein's Energy-Momentum Complex}
\vspace{6 pt}

It is interesting to calculate the total energy of the Taub cosmology
using several different measures in order to make a comparison of the results.
We begin with the Einstein prescription, for which the
energy-momentum complex is defined as:

\begin{equation}
{{\theta}_i}^k = \frac{1}{16 \pi} {{H_i}^{kl}}_{,l}
\end{equation}

with the superpotentials ${H_i}^{kl}$ given by:

\begin{equation}
{H_i}^{kl} = \frac{g_{in}}{\sqrt{-g}}{[-g (g^{kn} g^{lm} - g^{ln} g^{km})]}_{,m}
\end{equation}

which possesses the antisymmetric property that:
\begin{equation}
{H_i}^{kl} = -{H_i}^{lk}
\end{equation}

Substituting into (7)  the metric components of the Taub solution (1-5)
we find that the only non-zero components of ${H_i}^{kl}$ are:
\begin{equation}
{H_0}^{01} = -{H_0}^{10} = - \frac{k^2}{8} \hspace{0.1cm} \cos(x)
\hspace{0.1cm} {\rm sech}^2(\frac{k \hspace{0.1cm} t + \beta}{2})
\end{equation}

Therefore, from (6) the only non-zero components of ${{\theta}_i}^k$ are:
\begin{eqnarray}
{\theta_0}^{0} &=& \frac{k^2}{128 \pi} \hspace{0.1cm} \sin(x) \hspace{0.1cm}
{\rm sech}^2(\frac{k \hspace{0.1cm} t + \beta}{2}) \\
{\theta_0}^{1} &=& -\frac{k^3}{128 \pi} \hspace{0.1cm} \cos(x)
\hspace{0.1cm} {\rm sech}^2(\frac{k \hspace{0.1cm} t +
\beta}{2})\hspace{0.1cm} \tanh(\frac{k \hspace{0.1cm} t + \beta}{2})
\end{eqnarray}

In general, the energy-momentum components can be found by the volume integral:
\begin{equation}
P_i = \int \int \int {\theta_i}^0 dx^1 dx^2 dx^3
\end{equation}

Thus, if we substitute the coordinate ranges (2) we can write the total energy as:

\begin{equation}
P_0 = \int_0^{\pi} \int_0^{4\pi} \int_0^{2\pi} {\theta_0}^0 dx \hspace{0.1 cm} dy \hspace{0.1 cm} dz
\end{equation}

Evaluating this integral for (10), we obtain an expression for the
total energy of the Taub solution:

\begin{equation}
E= P_0 = \frac{\pi k^2 \hspace{0.1 cm}}{8} \hspace{0.1cm} {\rm
sech}^2(\frac{k \hspace{0.1cm} t + \beta}{2})
\end{equation}

\vspace{1 cm} \noindent
 {\bf 4. Following Papapetrou's Procedure}

\vspace{0.5 cm} Unlike Einstein's prescription, the energy-momentum
complex of Papapetrou has the advantage of being symmetric in its
indices.  Hence it allows local conservation laws to be well-defined.

The Papapetrou energy-momentum complex is defined as:

\begin{equation}
{\Omega}^{ik} = \frac{1}{16 \pi} {{N}^{ikab}}_{,ab}
\end{equation}

\begin{equation}
P_i = \int \int \int {\Omega}^{i0} dx^1 dx^2 dx^3
\end{equation}

where the functions ${N}^{ikab}$ are given by:

\begin{equation}
{N}^{ikab} = \sqrt{-g}\hspace{0.1 cm}[g^{ik} {\eta}^{ab} - g^{ia} {\eta}^{kb} +
g^{ab} {\eta}^{ik} - g^{kb} {\eta}^{ia}]
\end{equation}

\vspace{0.5 cm}

and the $\eta^{ab}$ terms represent the components
of a Minkowski metric of signature $-2$.

Inserting the metric components of the Taub solution (1-5) into expression (17),
we obtain the following relevant non-zero components of ${N}^{ikab}$:

\begin{eqnarray}
N^{0011} &=& N^{1100} = - \sin(x) \hspace{0.1cm}
[4 + k^2 \hspace{0.1 cm}  {\rm sech}^2(\frac{k \hspace{0.1cm} t + \beta}{2})]\\
N^{0101} &=& N^{1010} = \sin(x) \hspace{0.1cm}
[4 + k^2 \hspace{0.1 cm}  {\rm sech}^2(\frac{k \hspace{0.1cm} t + \beta}{2})]
\end{eqnarray}

Substituting into (15), we find that the only non-vanishing
component of the energy density is:

\begin{equation}
{\Omega}^{00} = \frac{1}{16 \pi} \hspace{0.1cm} \sin(x) \hspace{0.1cm}
[4 + k^2 \hspace{0.1 cm}  {\rm sech}^2(\frac{k \hspace{0.1cm} t + \beta}{2})]
\end{equation}

\vspace{0.5 cm}

To obtain the total energy, we integrate the energy density ${\Omega}^{00}$ over the appropriate coordinate ranges (2):

\begin{equation}
P_0 = \int_0^{\pi} \int_0^{4\pi} \int_0^{2\pi} {\Omega}^{00} dx \hspace{0.1 cm} dy \hspace{0.1 cm} dz
\end{equation}

This yields:

\begin{equation}
E= P_0 = \pi \hspace{0.1 cm} [4 + k^2 \hspace{0.1 cm}
 {\rm sech}^2(\frac{k \hspace{0.1cm} t + \beta}{2})]
\end{equation}

\vspace{1 cm} \noindent
 {\bf 5. Using the Technique of Landau and Lifshitz}

\vspace{0.5 cm}
Like Papapetrou's procedure, the Landau-Lifshitz
energy-momentum complex is symmetric in its indices, permitting
conservation laws, such as conservation of angular
momentum, to be readily defined.

We define the complex by means of the functions:

\begin{equation}
L^{ij} = \frac{1}{16 \pi} S_{,kl}^{ijkl}
\end{equation}

where the $L^{i0}$ represent the energy and momentum densities and:

\begin{equation}
S^{ijkl} = -g \hspace{0.1 cm} [g^{ij} g^{kl} - g^{ik} g^{jl}]
\end{equation}

Substituting into this expression the metric components of the Taub
solution (1-5) we find the only relevant non-zero component to be:

\begin{equation}
S^{0011} =  -\frac{k^2}{4} \hspace{0.1cm} \sin^2 (x) \hspace{0.1cm}
{\rm sech}^2(\frac{k \hspace{0.1cm} t + \beta}{2})
\end{equation}

From (23), we determine the energy density $L^{00}$ to be:

\begin{equation}
L^{00} =  -\frac{k^2}{32 \pi} \hspace{0.1cm} \cos (2x) \hspace{0.1cm}
{\rm sech}^2(\frac{k \hspace{0.1cm} t + \beta}{2})
\end{equation}

We now integrate this expression over the full range of coordinates:

\begin{equation}
P_0 = \int_0^{\pi} \int_0^{4\pi} \int_0^{2\pi} L^{00} \hspace{0.1 cm}
 dx \hspace{0.1 cm} dy \hspace{0.1 cm} dz
\end{equation}

Evaluating this integral, we find the total energy to be:

\begin{equation}
E= P_0 = 0
\end{equation}

So far, we have seen three different results for the total energy,
from the Einstein, Papapetrou and Landau-Lifshitz complexes respectively.
Perhaps this discrepancy stems from the coordinate system used, which is not
quasi-Cartesian.  To obtain coordinate-independent results, we now turn to M{\o}ller's
energy-momentum complex, specially designed to avoid such issues.

\vspace{1 cm} \noindent
 {\bf 6. Using M{\o}ller's Method}

\vspace{0.5 cm} M{\o}ller constructed his complex to transform spatially like a tensor
and hence be as free as possible of the choice of coordinate system.  Hence, unlike other
complexes, it does not favor quasi-Cartesian coordinates.  Given that the natural
coordinates to define the Taub solution are not quasi-Cartesian,  M{\o}ller's procedure
seems an ideal method to evaluate its gravitational energy.

The quantity M{\o}ller developed is defined as:

\begin{equation}
{\Xi}_i^k = \frac{1}{8 \pi} {\chi}_{i, p}^{kp}
\end{equation}

where:

\begin{equation}
{\chi}_i^{kl} = \sqrt{-g} \hspace{0.1 cm} [g_{ip, q} - g_{iq, p}] \hspace{0.1 cm}
g^{kq} \hspace{0.1 cm} g^{lp}
\end{equation}

Inserting the Taub metric components (1-5) we find that for all values of $k$ and $l$:

\begin{equation}
{\chi}_0^{kl} = 0
\end{equation}

Therefore, for all values of $k$:
\begin{equation}
{\Xi}_0^k = 0
\end{equation}

Integrating, we obtain the total energy:
\begin{equation}
E= \int_0^{\pi} \int_0^{4\pi} \int_0^{2\pi} {\Xi}_0^{0} \hspace{0.1 cm}
 dx \hspace{0.1 cm} dy \hspace{0.1 cm} dz = 0
\end{equation}

Thus,  M{\o}ller's energy-momentum complex agrees with
the prescription of Landau-Lifshitz that the total energy of
the vacuum Taub cosmology is zero.  We also find that all the momentum
components are zero.

\vspace{1 cm} \noindent
 {\bf 7. Conclusion}

\vspace{0.5 cm} \noindent We have investigated the total gravitational
energy of the vacuum-filled Taub cosmology using four different
energy-momentum complexes:  those of Einstein, Papapetrou,
Landau-Lifshitz and M{\o}ller.  Of these, the latter two yielded a value of
zero for the total energy of the universe, agreeing with Rosen's conjecture
that the total energy of any closed cosmology is zero.  The Einstein and
Papapetrou complexes, in contrast, yielded non-zero quantities that evolve
over time.  In our opinion, these two results are suspect because the natural
coordinates for the Taub metric, though holonomic, are not quasi-Cartesian,
and thus do not fulfill the ideal conditions for these complexes.

We note that our results from the M{\o}ller and Landau-Lifshitz complexes agree
with earlier work on anisotropic cosmologies, including evaluations of the
total energy of the Kasner (Bianchi type I) cosmology for several different cases.
We conjecture that the total energy of all vacuum-filled anisotropic
cosmologies is zero.  Further calculations, including other Bianchi types,
would be needed to confirm this hypothesis.

\vspace{1 cm} \noindent
 {\bf Acknowledgements}

\vspace{6 pt} \noindent Thanks to R. T. Jantzen and K. S. Virbhadra
for their valuable suggestions.  We also wish to thank Timothy Lucas
for his assistance.

\newpage

\vspace{1 cm} \noindent\fontsize{9 pt}{11 pt} \selectfont

\end{document}